\documentclass[useAMS,usenatbib]{mn2e}
\usepackage{times}

\title[The final lens candidate]{The final candidate from the JVAS/CLASS search for 6 arcsec to 15 arcsec image separation lensing}
\author[J. P. McKean]{J. P. McKean\thanks{mckean@astron.nl}\\
ASTRON, Oude Hoogeveensedijk 4, 7991 PD Dwingeloo, the Netherlands}
\begin{document}

\date{Accepted 2010 October 29. Received 2010 October 18; in original form 2010 October 5}

\pagerange{\pageref{firstpage}--\pageref{lastpage}} \pubyear{2010}

\maketitle

\label{firstpage}

\begin{abstract}
A search for 6 arcsec to 15 arcsec image separation lensing in the Jodrell Bank--Very Large Array Astrometric Survey (JVAS) and the Cosmic Lens All-Sky Survey (CLASS) by Phillips et al. found thirteen group and cluster gravitational lens candidates. Through radio and optical imaging and spectroscopy, Phillips et al. ruled out the lensing hypothesis for twelve of the candidates. In this paper, new optical imaging and spectroscopy of J0122+427, the final lens candidate from the JVAS/CLASS 6 arcsec to 15 arcsec image separation lens search, are presented. This system is found not to be a gravitational lens, but is just two radio-loud active galactic nuclei that are separated by $\sim$10~arcsec on the sky and are at different redshifts. Therefore, it is concluded that there are no gravitational lenses in the JVAS and CLASS surveys with image separations between 6 arcsec to 15 arcsec. This result is consistent with the expectation that group- and cluster-scale dark matter haloes are inefficient lenses due to their relatively flat inner density profiles.
\end{abstract}


\section{Introduction}

The Jodrell Bank--Very Large Array (VLA) Astrometric Survey (JVAS; \citealt{patnaik92,browne98,wilkinson98}) and the Cosmic Lens All-Sky Survey (CLASS; \citealt{browne03,myers03}) are the largest, statistically complete searches for gravitationally lensed radio-loud active galactic nuclei (AGN). Together, these two surveys found 22 gravitational lens systems with image separations between 0.3~arcsec~$\leq \Delta\theta_{\rm sep} \leq$~6~arcsec, which corresponds to the galaxy-scale regime of gravitational lensing. The lens systems discovered by JVAS and CLASS have been used, for example, to determine the mass profile of lensing galaxies \citep{cohn01,wucknitz04,suyu09}, investigate the impact of baryon cooling \citep{kochanek01}, quantify the level of low mass-substructure in the lensing haloes \citep{bradac02,biggs04,dalal02,kochanek04,mckean07b,jackson10}, measure the Hubble constant \citep{biggs99,koopmans00,fassnacht02,york05,suyu10} and investigate the high redshift Universe \citep{barvainis02,impellizzeri08}. The gravitational lensing statistics from the JVAS and CLASS surveys have also been used to constrain cosmological models \citep{chae02}.

The JVAS and CLASS surveys targeted flat-spectrum radio sources ($\alpha \geq -$0.5 where $S_{\nu} \propto \nu^{\alpha}$), which typically show compact radio emission on sub-arcsec scales. The surveys were carried out with the VLA at 8.46 GHz in A-configuration, which had a beam-size of $\sim$~200~mas and reached a sensitivity of $\sim$~200~$\mu$Jy in just 30-s of integration. During the course of JVAS and CLASS, over $\sim$15\,000 radio sources were observed with the VLA. However, using the NVSS (National Radio Astronomy Observatory VLA Sky Survey; \citealt{condon98}) catalogue at 1.4 GHz and the GB6 (Greenbank 6~cm; \citealt{gregory96}) catalogue at 4.85 GHz, a complete sample of 11\,685 flat-spectrum radio sources with $S_{\rm 4.85~GHz} \geq\,$30~mJy was defined. The observing set-up used for the JVAS and CLASS surveys made it possible to also search for examples of gravitational lensing by groups of galaxies and galaxy clusters. Lensing by more massive haloes like these would result in a larger separation between the lensed images. \cite{phillips01} carried out such a search by inspecting all of the images produced during the JVAS and CLASS VLA observations and found thirteen gravitational lensing candidates with image separations between 6~arcsec~$\leq \Delta\theta_{\rm sep} \leq$~15~arcsec. The upper-limit of this lens search was set to reduce the number of random alignments and limit the possibility of variability, coupled with a large lensing time-delay, changing the observed properties of the candidate lensed images.

These lens candidates were re-observed using the VLA and observed with MERLIN (Multi-Element Radio Linked Interferometric Network) between 1.4 and 15 GHz to investigate the radio spectral energy distributions, polarization and morphology of the radio sources. Optical spectroscopy was also carried out to determine the redshifts, in some cases. From these observations 12 of the candidates were rejected as gravitational lens systems with wide image separations. The one remaining candidate, J0122+427, has two radio sources (A and B) separated by 9.9~arcsec that are unresolved when observed with the VLA and MERLIN. Both radio sources have similar radio spectra up to 15 GHz, although there is evidence of a turnover in J0122+427B at around $\sim$~8.46~GHz. The flux-ratio (A$/$B) of the two radio sources is $\sim$~1.4 between 1.4 and 8.46 GHz, and increases to $\sim$~2.4 at 15 GHz \citep{phillips01}. Based on the radio properties of these two sources it was not possible to rule out the gravitational lensing hypothesis for this candidate.

To determine if the two radio sources from J0122+427 are gravitationally lensed or not, new optical imaging and spectroscopy of the system has been carried out, which are presented in this paper. The observations and results are presented in Sections \ref{obs} and \ref{results}, respectively. The lensing hypothesis for this system and gravitational lensing by galaxy groups are discussed in Section \ref{disc+conc}.

\section{Observations \& Data Analysis}
\label{obs}

Optical imaging and spectroscopic observations of J0122+427A$/$B were carried out with the 4.2-m William Herschel Telescope (WHT) at the Observatorio del Roque de los Muchachos on La Palma. These observations were taken on 2000 September 25 and 26 as part of the WHT service programme. The conditions on both nights were clear and the full width at half maximum (FWHM) seeing was $\sim$0.5--0.7~arcsec.

The imaging data were taken through the Aux-port imager, which has an approximate field-of-view of 2 arcmin and has 0.216-arcsec sized pixels. Single 300-s exposures of the J0122+427 field were taken with the {\it V-} and {\it I}-band filters. The faint standard star SA--92--426 \citep{landolt92} was also observed to determine the flux calibration. The astrometric calibration was added to the images by fitting to the positions of United States Naval Observatory (USNO) stars; the resulting fit had an rms of $\sim$0.35~arcsec, which includes the systematic scatter in the USNO catalogue. 

The spectroscopic observations were carried out using the Intermediate-dispersion Spectrograph and Imaging System (ISIS). This instrument has a blue and a red arm that can be used simultaneously to obtain data over the optical part of the spectrum. The R158B (3.25~{\AA}~pixel$^{-1}$; centred at $\lambda_{\rm c}=$~4606~\AA) and R158R (2.91~{\AA}~pixel$^{-1}$; centred at $\lambda_{\rm c} =$~7599~{\AA}) gratings were used with the blue and red arms, respectively, and together provided spectra between 3000 and 9300 {\AA}. The spectra were taken using a 1.5~arcsec wide longslit and as 3$\,\times\,$1800-s exposures, giving a total integration time of 1.5 h. Exposures of CuAr and CuNe arc-lamps were taken either before or after each set of science observations to determine the wavelength calibration, which had an rms of $\sim$0.3~{\AA}. The standard star Feige~110 \citep{oke90} was observed using the same set-up as the science frames to remove the response of the spectrograph and to obtain an approximate flux calibration. The imaging and spectroscopic data were reduced within {\sc iraf} (Image Reduction and Analysis Facility) using standard procedures (see \citealt{mckean04} for details).

\section{Results}
\label{results}

\begin{figure*}
\begin{center}
\setlength{\unitlength}{1cm}
\begin{picture}(12,7)
\put(-1.2,0){\includegraphics{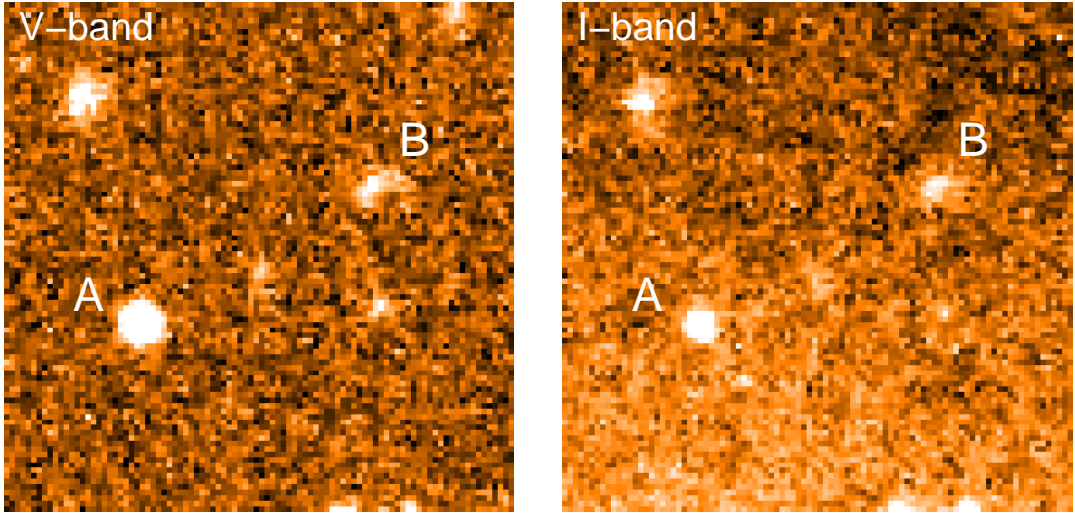}}
\end{picture}
\caption{The optical imaging of the candidate gravitationally lensed images at {\it V}- and {\it I}-bands taken with the WHT. The images show a sky-area of 20\,$\times$\,20~arcsec$^{2}$ around the mid-point between the two radio sources. North is up and east is to the left.} 
\label{imaging}
\end{center}
\end{figure*}

The {\it V}- and {\it I-}band images of the candidate gravitational lens system are shown in Fig.~\ref{imaging}. The optical counterpart of J0122+427A is detected 0.7 arcsec from the radio position and is found to be a point source at both bands. There is a faint and extended galaxy which is 0.9 arcsec from the expected position of J0122+427B, which is assumed to be the optical counterpart of the radio source. Photometry of the two optical identifications was carried out using circular apertures with 5-arcsec diameters and the apparent magnitudes are given in Table~\ref{photometry}. The flux-ratios (A$/$B) of the two sources at {\it V}- and {\it I-}band are $\sim$11 and $\sim$6, respectively. Note that if the faint extended galaxy is not associated with J0122+427B, then based on the 3$\sigma$ limiting magnitudes, the flux-ratio between A and B would be $\ga$\,30 and $\ga$\,10 in the {\it V}- and {\it I-}bands, respectively. There is no evidence of a cluster or group in the form of luminous red galaxies around the lens system, although a very faint object is found between the two candidate lensed images.

\begin{figure*}
\begin{center}
\setlength{\unitlength}{1cm}
\begin{picture}(12,6.2)
\put(-3.3,0){\includegraphics{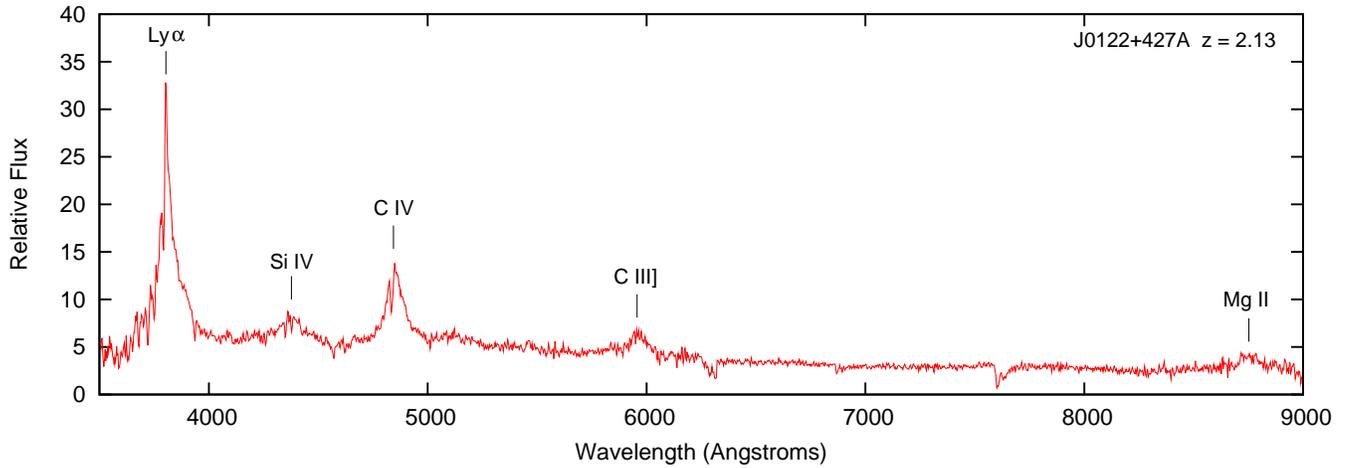}}
\end{picture}
\caption{The optical spectrum of J0122+427A taken with the WHT. The spectrum shows strong continuum emission and five emission lines at redshift 2.130\,$\pm\,$0.002. The relative flux-scale of the spectrum is in units of 10$^{-17}$~erg~cm$^{-2}$~s$^{-1}$~{\AA}$^{-1}$.} 
\label{spectra-a}
\end{center}
\end{figure*}

\begin{figure*}
\begin{center}
\setlength{\unitlength}{1cm}
\begin{picture}(12,7.7)
\put(-3.3,0){\includegraphics{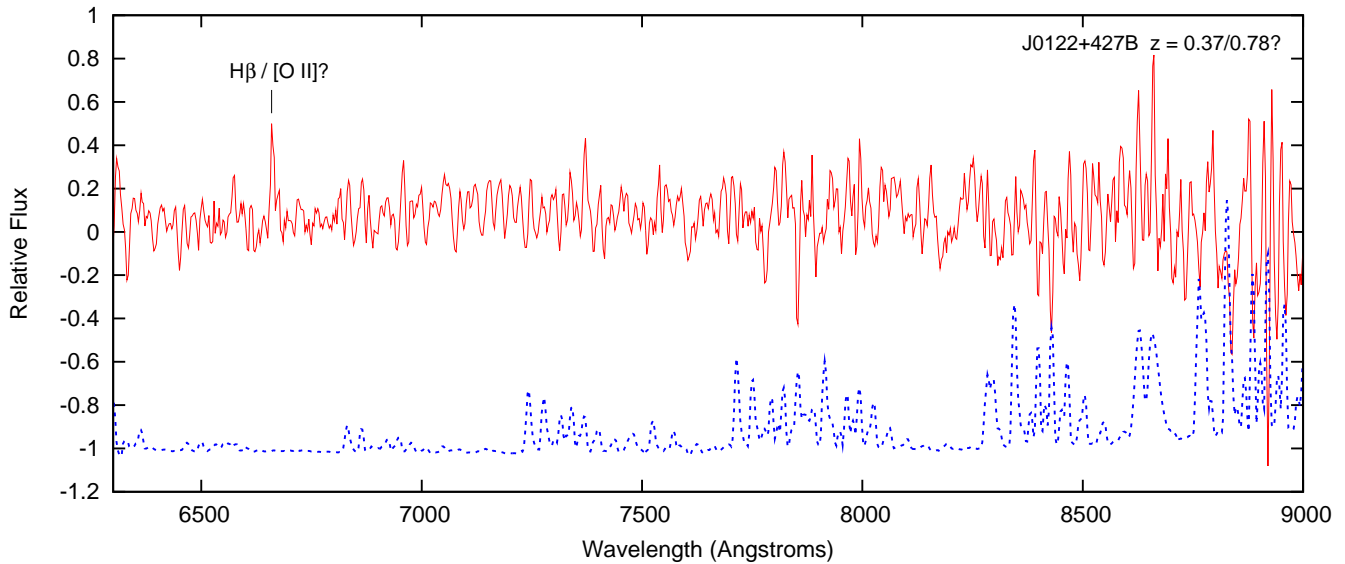}}
\end{picture}
\caption{The optical spectrum of J0122+427B taken with the WHT (solid; red) and the scaled sky spectrum (dashed; blue). Only faint continuum emission from the radio source is detected. There is a single emission line found at 6661~{\AA} that is not coincident with a sky line, a sky absorption band or a cosmic ray event. Assuming this line is either H$\beta$ or [O~{\sc ii}], then the redshift of J0122+427B is 0.37 or 0.78, respectively. The relative flux-scale of the spectrum is in units of 10$^{-17}$~erg~cm$^{-2}$~s$^{-1}$~{\AA}$^{-1}$.} 
\label{spectra-b}
\end{center}
\end{figure*}

The optical spectra of the two radio sources are shown in Figs.~\ref{spectra-a} and \ref{spectra-b}. The spectrum of J0122+427A has a strong continuum flux and shows five broad emission lines that are identified as Ly${\alpha}$, Si\,{\sc iv}, C\,{\sc iv}, C\,{\sc iii}  and Mg\,{\sc ii} at redshift 2.13. The results of Gaussian profile fits to these lines are given in Table \ref{lines}. The high redshift, broad emission lines, blueish colour and point-like optical morphology are all consistent with J0122+427A being a quasar. Interestingly, the two highest signal-to-noise ratio lines in the spectrum, Ly${\alpha}$ and C\,{\sc iv}, also show blue-shifted absorption from presumably out-flowing gas. The spectrum of J0122+427B is much weaker and the continuum was only marginally detected in the red part of the spectrum, consistent with the red colour found from the optical imaging. There is a 7$\sigma$ detection of an emission line at 6661~{\AA}, which is not coincident with a sky-line or a cosmic ray event. However, this line does not correspond to anything at redshift 2.13. Likely identifications for the line are [O\,{\sc ii}] at redshift 0.78 or H$\beta$ at redshift 0.37, given the faintness of the optical emission, which probably rules out H$\alpha$ at very low redshift. The detection in the {\it V}-band imaging probably rules out Ly$\alpha$ at high redshift because the flux blue-ward of the line would be absorbed away as part of the Lyman forrest.

\begin{table*}
\begin{center}
\caption{Results from the optical photometry and spectroscopy of the two candidate gravitationally lensed radio sources. The photometry has been carried out using a circular aperture with a 5 arcsec diameter. The positions are taken from the CLASS 8.46 GHz radio observations \citep{myers03} and are in J2000 co-ordinates. The offsets to the optical positions are given relative to the radio positions.}
\begin{tabular}{lcccccccl} \hline
Image			& RA				& Dec.					& $\Delta$RA		& $\Delta$Dec.		& {\it V}					& {\it I}			& $V-I$	& Redshift	\\
				& ($^{h~m~s}$)		& ($\degr~\arcmin~\arcsec$)	& (arcsec)			& (arcsec)			& \multicolumn{3}{c}{(Vega magnitudes)}					&					\\ \hline
J0122+427A	&	01:22:24.735		& +42:45:21.10				& $+$0.61			& $+$0.26			& 20.53\,$\pm$\,0.02	& 19.96\,$\pm$\,0.05	& 0.57			& 2.130\,$\pm\,$0.002	\\
J0122+427B	&	01:22:24.026		& +42:45:27.19				& $-$0.79			& $-$0.46			& 23.12\,$\pm$\,0.21	& 21.82\,$\pm$\,0.25	& 1.30			& 0.37$/$0.78?			\\ \hline
\end{tabular}
\label{photometry}
\end{center}
\end{table*}

\begin{table}
\begin{center}
\caption{The wavelengths of the emission lines detected from J0122+427A at redshift 2.130\,$\pm\,$0.002.}
\begin{tabular}{lcccc} \hline
Line			& $\lambda_{\rm rest}$	& $\lambda_{\rm obs}$	& Redshift \\
			& (\AA)				& (\AA)				&		 \\ \hline
Ly$\alpha$	& 1216				& 3811				& 2.134	 \\
Si {\sc iv}		& 1398				& 4377				& 2.131	 \\
C {\sc iv}		& 1549				& 4854				& 2.134	 \\
C {\sc iii}]		& 1909				& 5962				& 2.123	 \\
Mg {\sc ii}		& 2798				& 8752				& 2.127	 \\ \hline
\end{tabular}
\label{lines}
\end{center}
\end{table}

\section{Discussion \& Summary}
\label{disc+conc}

The optical imaging and spectroscopic data for J0122+427A$/$B rules out the lensing hypothesis for this system. This conclusion is based on the differences in the optical morphologies, the inconsistent optical and radio flux-ratios, the lack of a suitable lens between the two candidate images and from their differing spectra and redshifts. Therefore, the JVAS/CLASS search for 6~arcsec~$\leq \Delta\theta_{\rm sep} \leq$~15~arcsec lensing has found no new gravitational lens systems.

The lensing statistics of the JVAS and CLASS 6~arcsec to 15~arcsec image separation lens search were analyzed by \citet{phillips01}. They concluded that if the lensing mass distributions of groups and clusters were isothermal, as is the case for galaxy-scale lenses (e.g. \citealt{koopmans09}), then there should be around four new lens systems found from the JVAS/CLASS complete sample of 11\,685 flat-spectrum radio sources. The lack of gravitational lensing in this range of image separations was interpreted as galaxy groups and clusters being less efficient lenses due to a flattening of their total dark and luminous mass profiles (see \citealt{phillips01} for a detailed discussion), in agreement with the results from numerical simulations (e.g. \citealt{navarro96}) and lensing studies of clusters (e.g. \citealt{sand02}). 

The largest image separation lens system found by the JVAS and CLASS surveys is B2108+213 \citep{mckean05,fassnacht08,more08}, which has two lensed images of a radio-loud AGN that are separated by 4.56 arcsec. Further observations found that the B2108+213 lens is a group of galaxies and that the inner mass profile of the lens is consistent with an isothermal mass distribution \citep{mckean10}. However, in this case the lensing is dominated by the brightest group galaxy. This galaxy is massive and is at the centre of the group, which increases the local surface-mass density above the critical limit for lensing. Therefore, the case of B2108+213 may not be indicative of a typical galaxy group and it may be the case that group-scale mass distributions are in general flatter than their galaxy-scale counterparts.

The rarity of group and cluster lensing of AGN is highlighted by the few confirmed cases of gravitational lenses with large lensed-image separations. There are only three known gravitationally lensed AGN that have image separations larger than 6 arcsec; B0957+561, the first gravitational lens discovered \citep*{walsh79} and two systems from the Sloan Digital Sky Survey (SDSS), namely, J1004+4112 \citep{inada03} and J1029+2623 \citep{inada06}. Based on the gravitational lensing statistics of the SDSS, the probability of finding a wide-image separation gravitational lens system is 1:36287 \citep{inada10}. This is much smaller than the rate of finding galaxy-scale lenses, for example, 1:2016 for the SDSS \citep{inada10} and 1:689 for the JVAS/CLASS \citep{chae02} lens searches. Note that the selection function of the SDSS and the JVAS/CLASS surveys are different, which is why their lensing probabilities are not the same. 

To find additional cases of group and cluster lensing will require a larger number of the background source population to be surveyed. Based on the number counts of flat-spectrum radio sources \citep{mckean07a}, extending the JVAS and CLASS surveys down to $S_{\rm 4.85~GHz} \geq $~5~mJy will provide a parent population of $\sim$70\,000 radio sources. However, a sample this large is only expected to increase the number of wide-image separation lens systems known by a few, assuming that the redshift distribution of these sources is similar to that of the JVAS and CLASS parent samples. Therefore, it seems that searches for gravitationally lensed AGN will not be worthwhile when it comes to looking for group and cluster gravitational lensing at radio wavelengths in the future. Alternatively, the number counts of star-forming galaxies is significantly larger than the radio-loud AGN population and offer a much better chance to observe gravitational lensing by groups and clusters. Indeed, almost all of the cluster lenses that are currently known have star-forming galaxies as the lensed source, that are selected by their optical or sub-mm emission. Recently, there has also been a large increase in the number of potential group-scale lenses discovered that show extended gravitational arcs from background blue star-forming galaxies with image separations between 6 and 16 arcsec \citep{limousin09}. The prospects of carrying out surveys for radio emitting star-forming galaxies behind clusters has become more appealing with the advent of new or upgraded radio facilities (see previous work by \citealt{garrett05} and \citealt{berciano-alba10}). Simulations by \citet{fedeli09} find that over the whole sky, several 100 cluster lens systems could potentially be found from deep radio observations ($\sim$20~$\mu$Jy~arcsec$^{-2}$ surface-brightness sensitivity at 1.4~GHz) that are within the sensitivity limits of, for example, the Expanded VLA.

\section*{Acknowledgments}
JPM would like to thank Ian Browne and Neal Jackson for useful discussions and Paul Phillips for providing the WHT data. The William Herschel Telescope and its service programme are operated on the island of La Palma by the Isaac Newton Group in the Spanish Observatorio del Roque de los Muchachos of the Instituto de Astrofisica de Canarias.

\bsp

\label{lastpage}

\end{document}